\documentclass[twocolumn]{aastex63}

\shorttitle{X-ray luminosity dependent CL-AGN: UGC\,3223}
\shortauthors{Wang et al.}
\begin{document}

\title{An X-ray Luminosity-dependent ``Changing-look'' Phenomenon in UGC\,3223}

\correspondingauthor{J. Wang \& D. W. Xu}
\email{wj@nao.cas.cn, dwxu@nao.cas.cn}

\author{J. Wang}
\affiliation{Guangxi Key Laboratory for Relativistic Astrophysics, School of Physical Science and Technology, Guangxi University, Nanning 530004,
Peopleʼs Republic of China}
\affiliation{Key Laboratory of Space Astronomy and Technology, National Astronomical Observatories, Chinese Academy of Sciences, Beijing 100101,
Peopleʼs Republic of China}

\author{D. W. Xu}
\affiliation{Key Laboratory of Space Astronomy and Technology, National Astronomical Observatories, Chinese Academy of Sciences, Beijing 100101,
Peopleʼs Republic of China}
\affiliation{School of Astronomy and Space Science, University of Chinese Academy of Sciences, Beijing, Peopleʼs Republic of China}

%\collaboration{1}{(AAS Journals Data Scientists collaboration)}

%\author{S. Komossa}
%\affiliation{Max-Planck-Institut f\"ur Radioastronomie, Auf dem H\"ugel 69, D-53121 Bonn, Germany}

%\nocollaboration{1}

\author{J. Y. Wei}
\affiliation{Key Laboratory of Space Astronomy and Technology, National Astronomical Observatories, Chinese Academy of Sciences, Beijing 100101,
Peopleʼs Republic of China}
\affiliation{School of Astronomy and Space Science, University of Chinese Academy of Sciences, Beijing, Peopleʼs Republic of China}




\begin{abstract}

The nature of the rare ``Changing-look'' (CL) phenomenon in active galactic nuclei (AGNs) is still under debate at current stage.
We here present \it Swift/\rm XRT and UVOT follow-up observations of UGC\,3223, a newly discovered repeat CL-AGN with type transitions of
$\mathrm{S1.5\rightarrow S2 \rightarrow S1.8}$ occurring in a period of about 30 years. By comparing the values previously reported in the \it ROSAT \rm All-sky
Survey and in the second Swift-XRT Point Source catalog, we clearly find that the X-ray flux tightly follows the optical spectral transition, in which a 
spectral type closer to a Seyfert 1 type is associated with a higher X-ray flux. An invariable X-ray spectral shape is, however, found in the 
CL phenomenon of the object. An extremely low Eddington ratio of $\sim2\times10^{-4}$ can be obtained from the X-ray luminosity for its 
Seyfert 2 state, which suggests a favor of the disk-wind broad-line region model in explaining the CL phenomenon.
A variation of the total UV emission is not revealed when compared to the previous \it GALEX \rm NUV observation, since
the UVOT images indicate that 
$\sim90$\% UV emission comes from the intensive star formation in the host galaxy. 

\end{abstract}

\keywords{galaxies: Seyfert --- galaxies: nuclei --- X-rays: galaxies --- ultraviolet: galaxies}


\section{Introduction} \label{sec:intro}

A present hot topic is the nature of the rare ``changing-look'' active galactic nuclei (CL-AGNs)
that manifest themselves a spectral type transition between type I, intermediate type, and type II within a timescale of an order of
years to decades. Up to now, there are only $\sim100$ CL-AGNs identified by multi-epoch spectroscopy
(e.g., MacLeod et al. 2010, 2016; Shapovalova et al. 2010; Shappee et al. 2014;
LaMassa et al. 2015; McElroy et al. 2016; Ruan et al. 2016;
Runnoe et al. 2016; Parker et al. 2016; Gezari et al. 2017; Sheng et al. 2017, 2020; Stern et al. 2018; Yang et al. 2018; Wang et al. 2018, 2019, 2020;
Frederick et al. 2019; Trakhtenbrot et al. 2019; Yan et al. 2019; Ai et al. 2020; Graham et al. 2020; Kollatschny et al. 2018, 2020).
Although almost all the CL-AGNs are identified by a dramatic variation of their Balmer emission lines, Guo et al. (2019)
recently reported the first \ion{Mg}{2} CL-AGN: SDSS\,J152533.60+292012.1.   
The discovery of the CL-AGNs challenges the widely accepted AGN paradigm, not only in the orientation-based unified model (e.g., Antonucci 1993), but
also in the standard disk model in terms of the viscosity crisis (e.g., Lawrence 2018 and references therein).

The physical origin of CL-AGNs is still under debate, although some progress has been achieved in past a couple of years.
Based on the light echo in middle infrared (Sheng et al. 2017) and spectropolarimetry (e.g., Hutsemekers et al. 2019),
there is evidence suggesting that a variation
of the accretion power of a supermassive black hole (SMBH) is a plausible explanation of the rare CL phenomenon.
The statistical study in MacLeod et al. (2019) and case study in Wang et al. (2020) recently argue that the CL phenomenon can be
possibly understood in the context of the disc-wind broad-line region (BLR) model (e.g., Elitzur \& Ho 2009; Nicastro 2000; Elitzur \& Shlosman 2006),
in which a critical Eddington ratio ($L/L_{\mathrm{Edd}}$, $L_{\mathrm{Edd}}=1.26\times10^{38}M_{\mathrm{BH}}/M_\odot\ \mathrm{erg\ s^{-1}}$ is 
the Eddington luminosity) of $\sim10^{-3}$ ($10^{-6}$) is predicted for an (dis)appearance of broad emission lines.

Ruan et al. (2019) recently found an anti-correlation between UV-to-X-ray spectral index and $L/L_{\mathrm{Edd}}$ in some CL-AGNs, which suggests 
a linkage between the accretion state transition occurring in X-ray binaries (XRBs) and the CL phenomenon observed in AGNs. This linkage was subsequently 
reinforced by Ai et al. (2020 and references therein) who identified a relation between spectral type transition and variation of X-ray hardness in
a small sample of CL-AGNs, i.e., a well-established V-shape correlation between X-ray hardness and $L/L_{\mathrm{Edd}}$.

Wang et al. (2020) recently identified a new nearby CL-AGN, UGC\,3223 (1RXS\,J045909.6+045831, $z=0.015621$), with a repeat type transition. 
So far, repeat type transitions have been only identified in a few nearby Seyfert galaxies and quasars (e.g., Parker et al. 2019; 
Marin et al. 2019 and references therein). Based on the spectrum taken by Stirpe (1990) in 1987,
our spectroscopic observations of the object over the course of 18 yr enable us to witness its type transition from
1.5$\rightarrow$2.0$\rightarrow$1.8 over 32 yr. Our latest spectroscopy in 2020/02 suggests the object is now still 
in the ``turn-on'' state with a Seyfert 1.8 spectrum.

In this paper, we report \it Swift/\rm XRT and UVOT follow-up observations, taken at the ``turn-on'' state, of this interesting object, and 
explore the origin of its CL phenomenon by examining the variation of its X-ray and UV emission by comparing with previous results extracted 
from literature. The paper is organized as follows. 
Section 2 describes \it Swift/\rm XRT and UVOT follow-up observations and data reductions.
The variation of its X-ray and UV emission is studied in Section 3. Section 4 shows the conclusion and discussion. A $\Lambda$CDM cosmology with
parameters $H_0=70 \mathrm{km\ s^{-1}\ Mpc^{-1}}$, $\Omega_m=0.3$ and $\Omega_{\Lambda}=0.7$
is adopted throughout the paper.

\section{Observations and Data Reductions} \label{sec:style}

We proposed X-ray and ultraviolet follow-up observations for the object in 2020 by using the Neil Gehrels \it Swift \rm Observatory 
(Gehrels et al. 2004) X-ray telescope (XRT) and Ultraviolet/Optical Telescope (UVOT) . 
The object was targeted three times (ObsID=00049202002, 00049202003 and 00049202004) by XRT and UVOT simultaneously on 2020 March 31, April 6 and April 10.     
The log of these observations is shown in Table 1.

\subsection{XRT X-ray Spectrum }
We reduced the XRT data taken in the Photon Counting (PC) mode by the HEASOFT version 6.27.2, along with the corresponding CALDB version 20190910.
All the three XRT exposures were at first stacked to enhance the signal-to-noise (S/N) ratio.
The source spectrum was then extracted from the stacked image by a circular region with a radius of 23.5\arcsec.
The corresponding background spectrum was extracted from an adjacent region free of any sources.
The corresponding ancillary response file was produced by the task \it xrtmkarf. \rm
In fact, the XRT count rates in 0.3-10keV are $(4.77\pm0.75)\times10^{-2}$, $(4.51\pm1.37)\times10^{-2}$ and $(5.14\pm0.80)\times10^{-2}\ \mathrm{count\ s^{-1}}$
for the three exposures, which suggests a stable count rate level within the period.

The stacked X-ray spectrum was then modeled by XSPEC (v12.11, Arnaud 1996) in terms of four models. 
The best fitting and best-fit parameters are shown in Figure 1 and Table 2, respectively. 
All the uncertainties quoted in the table correspond to a 90\% significance level.

%Because the count rates are not high enough, we
We first modeled the spectrum by a simple model expressed as $wabs*powerlaw$ over the 0.3-10keV in terms of the 
C-statistic (Cash 1979; Humphrey et al. 2009; Kaastra 2017), where the Galactic hydrogen column density 
is fixed to be $N_{\mathrm{H}}=8.1\times10^{20}\ \mathrm{cm^{-2}}$ (Kalberla et al. 2005). 
%The best fitting is reproduced in the upper panel in Figure 1 by the red line. 
%The best fitted parameters, along with the uncertainties, are listed in Table 2. 
%All the uncertainties correspond to a 90\% significance level. 
Although the fitting is acceptable with a C-statistic of 1.283, it returns a quite hard spectrum with a photon 
index of $\Gamma=0.82^{+0.29}_{-0.30}$. This unusually hard spectrum implies either an existence of strong
excess absorption, or else a significant starburst contribution, although the latter is unlikely (see Section 4).

In order to include a potential local absorption, a model of $wabs*zwabs*powerlaw$ was then used to reproduce the observed spectrum. 
However, the best-fit returns a zero value of the intrinsic hydrogen column density $N_\mathrm{H}$ when it is set as a free parameter in the modeling. A 
fixed value of $N_\mathrm{H}$ was therefore adopted in our modeling. 
With the fixed intrinsic $N_\mathrm{H}$, a softer powerlaw with $\Gamma=1.24^{+0.33}_{-0.34}$ can be obtained from the 
best fit. We determine the $N_\mathrm{H}$ from either \ion{H}{1} 21cm observation or extinction determined
from the Balmer decrement as follows. 

We estimate the intrinsic $N_\mathrm{H}$ from the total \ion{H}{1} mass $M_\mathrm{HI}$ as $N_\mathrm{H}=M_\mathrm{HI}/\pi ab$, where $a$ and $b$ are 
the major and minor radius of the host galaxy, respectively. 
Given the standard way of estimating $M_\mathrm{HI}$ from the observed \ion{H}{1} 21cm line flux (Roberts 1962), the 
intrinsic $N_\mathrm{H}$ can be calculated from 
\begin{equation}
  N_\mathrm{H}=\frac{1.218\times10^{24}}{a\arcsec b\arcsec}\frac{F'_{\mathrm{HI}}}{\mathrm{Jy\ km\ s^{-1}}}\ \mathrm{cm^{-2}}
%M_{\mathrm{HI}}=2.356\times10^5\bigg(\frac{D}{\mathrm{Mpc}}\bigg)^2\bigg(\frac{F^{c}_{\mathrm{HI}}}{\mathrm{Jy\ km\ s^{-1}}}\bigg)M_\odot
\end{equation}
where both $a\arcsec$ and $b\arcsec$ are in units of arcsecond, and $F'_{\mathrm{HI}}$ is related with the observed 
21cm line flux as $F'_{\mathrm{HI}}=(a/b)^{0.12}F_{\mathrm{HI}}$ (Giovanelli et al. 1994). 
Using the measured $F_{\mathrm{HI}}=3.05\ \mathrm{Jy\ km\ s^{-1}}$ (Courtois et al. 2009) enables us to estimate intrinsic
$N_\mathrm{H}$ to be $\sim1.3\times10^{21}\ \mathrm{cm^{-2}}$. We alternatively estimate $N_\mathrm{H}$ from
the standard relationship between $N_\mathrm{H}$ and extinction $A_V$:
$N_\mathrm{H}/A_V\sim(1.8-2.2)\times10^{21} \mathrm{atoms\ cm^{-2}\ mag^{-1}}$, where $A_V$ can be determined from the 
the observed Balmer decrement of the narrow emission lines. With the standard Case B recombination, 
an extinction of $A_V=1.08$mag can be inferred from the spectroscopy reported in Wang et al. (2020), which finally
leads to a $N_\mathrm{H}=1.9-2.4\times10^{21}\ \mathrm{cm^{-2}}$.

%With the fixed intrinsic $N_\mathrm{H}$, a softer powerlaw with $\Gamma=1.24^{+0.33}_{-0.34}$ can be obtained from the 
%best fit. 
%The best-fit is shown in the upper panel in Figure 1 by the blue line, and the results are 
%listed in Table 2.

The spectrum can be also reproduced by a model of $wabs*zpcfabs*powerlaw$ by including a
neutral partially covering absorber. This modeling results in a little better 
C-statistic of 0.868 and a photon index of $\Gamma=2.06^{+0.69}_{-1.09}$ that is close to 
the mean value of AGNs of $\sim1.8$.
%The best fitting and fitted parameters are again shown in Figure 1 (the red line in lower panel) and Table 2, respectively.

Taking into account the intensive star formation activity occurring in the host disk (see Section 2.2), 
the spectrum is found to be best reproduced by a model including an emission from hot, diffuse gas (Raymond \& Smith 1977), i.e.,
$wabs*zwabs*(powerlaw+raymond)$.  
In this case, we obtain a photon index of $\Gamma=2.13\pm0.66$ and a plasma electronic temperature of $kT_\mathrm{e}=0.05^{+0.02}_{-0.01}$ keV, 
along with a C-statistic of 1.023.

\subsection{UVOT Images}
 
The UVOT images were reduced by the HEASOFT and corresponding CALDB version 20170922. Again, we 
at first co-add the multi-exposures to enhance the S/N ratio for each filter. The corresponding exposure maps of the co-added images were 
generated by the \it uvotexpmap \rm task. Figure 2 displays the co-added UVOT images in the three used filters, i.e., $uvm2$, $uvw1$ and $uvw2$ from left to right.
The object is spatially resolved in all the UVOT filters. One can see from the three images that 
the UV emission is dominantly contributed from a ring and a spot in the host disk, rather than the nucleus, which indicates an intensive star formation activity 
occurring in the host disk. We argue that the revealed ring structure in UV with an extent of about 30\arcsec\  is most likely due to the large scale dust 
lane that was identified in the HST WFPC2 image by Deo et al. (2006). 

%In order to measure the total light from the object, 
An elliptical aperture with a radius of 36.4\arcsec, which is twice the effective radius 
reported in Bai et al. (2015), was adopted to measure its total brightness. Again, the corresponding background level was determined
from an adjacent region free of any sources.
The Galactic extinction is corrected by using the Galactic Reddening Map 
(Schlegel et al. 1998) and the Galactic extinction curve in Cardelli et al. (1989). The results of our photometry in the AB magnitude system are 
$uvm2=16.31\pm0.04$ mag, $uvw1=16.10\pm0.04$ mag and $uvw2=16.58\pm0.03$ mag,
where the reported magnitude errors include both statistical and system uncertainties.

\begin{table*}[h!]
\renewcommand{\thetable}{\arabic{table}}
\centering
%\tiny
\caption{Log of observations carried out by \it Swift\rm/XRT and UVOT.}
\label{tab:decimal}
\begin{tabular}{ccccc}
\tablewidth{0pt}
\hline
\hline
ObsID & Obs start time & XRT exposure in PC mode & UVOT exposure & UVOT filters\\
      &                & seconds &  seconds & \\
(1)  &   (2) & (3)  &(4) & (5)\\
\hline
00049202002 & 2020-03-31 21:19:34 & 857.5 &  845.1 & $uvm2,uvw1,uvw2$ \\
00049202003 & 2020-04-06 23:56:35 & 268.3 &  261.7 & $uvm2,uvw2$\dotfill \\
00049202004 & 2020-04-10 06:08:35 & 818.3 &  808.0 & $uvm2,uvw1,uvw2$\\
\hline
\hline
\end{tabular}
%\tablenotetext{a}{The limit magnitude at a significance level of $3\sigma$ is derived from the measurements of the field stars. }
\end{table*}

\begin{figure}[ht!]
\plotone{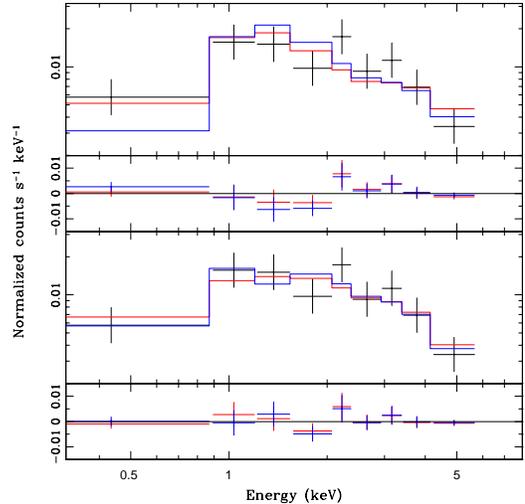}
\caption{\it Upper panels: Swift\rm/XRT X-ray spectrum of UGC\,3223
and the best-fit spectral model expressed as $wabs*powerlaw$ (Model 1, red line) and $wabs*zwabs*powerlaw$ (Model 2, blue line). 
The sub-panel underneath the spectrum is the 
deviations, in unit of $\mathrm{counts\ s^{-1}\ keV^{-1}}$, of the observed data from the best-fit model.
\it Lower panels: \rm the same as in the upper panel but for models of $wabs*zpcfabs*powerlaw$ (Model 3, red line) and 
$wabs*zwabs*(powerlaw+raymond)$ (Model 4, blue line).
\label{fig:general}}
\end{figure}

\begin{table*}[h!]
\renewcommand{\thetable}{\arabic{table}}
\centering
\caption{X-ray spectra fit parameters of \it Swift \rm XRT observations for UGC\,3223}
\label{tab:decimal}
\begin{tabular}{llll}
\tablewidth{0pt}
\hline
\hline
Parameter & Value & Units & Description \\
(1)  &   (2) & (3)  &(4)\\
\hline
\multicolumn{4}{l}{Model 1 - $wabs*powerlaw$}\\
%$n_{\mathrm{H}}$ & 0  & $10^{22}\mathrm{cm^{-2}}$ & Local column density\\
$\Gamma$         & $0.82^{+0.29}_{-0.30}$  &  & Powerlaw index\\
$F(\mathrm{2-10keV})$ & $3.44^{+1.38}_{-1.00}\times10^{-12}$ & $\mathrm{erg\ s^{-1}\ cm^{-2}}$ & Unabsored flux\\
Cash statistics  & 7.70/7=1.100 & & \\
\multicolumn{4}{l}{Model 2 - $wabs*zwabs*powerlaw$}\\
$N_{\mathrm{H}}$ & 0.24  & $10^{22}\mathrm{cm^{-2}}$ & Local column density (fixed)\\
$\Gamma$         & $1.24^{+0.33}_{-0.34}$  &  & Powerlaw index\\
$F(\mathrm{2-10keV})$ & $2.86^{+1.18}_{-0.85}\times10^{-12}$ & $\mathrm{erg\ s^{-1}\ cm^{-2}}$ & Unabsored flux\\
Cash statistics  & 10.77/7=1.539 & & \\
\multicolumn{4}{l}{Model 3 - $wabs*zpcfabs*powerlaw$}\\
$\eta_{\mathrm{H}}$ & $2.74_{-1.95}^{+2.03}$  & $10^{22}\mathrm{cm^{-2}}$ & Local equivalent column density\\
$f$                & $0.89_{-0.61}^{+0.09}$  &  & Dimensionless covering fraction\\
$\Gamma$           & $2.06^{+0.69}_{-1.09}$  &  & Powerlaw index\\     
$F(\mathrm{2-10keV})$ & $3.13^{+1.33}_{-0.93}\times10^{-12}$ & $\mathrm{erg\ s^{-1}\ cm^{-2}}$ & Unabsored flux\\
Cash statistics  & 4.23/5=0.868 & & \\

\multicolumn{4}{l}{Model 4 - $wabs*zwabs*(powerlaw+raymond)$}\\
$N_{\mathrm{H}}$ & $2.00^{+0.99}_{-0.58}$  & $10^{22}\mathrm{cm^{-2}}$ & Local column density (free)\\
$\Gamma$           & $2.13\pm0.66$  &  & Powerlaw index\\
$kT_\mathrm{e}$             & $0.05^{+0.02}_{-0.01}$ & keV & Temperature of hot, diffuse gas\\
$F(\mathrm{2-10keV})$ & $2.88^{+0.61}_{-0.53}\times10^{-12}$ & $\mathrm{erg\ s^{-1}\ cm^{-2}}$ & Unabsored flux of the powerlaw\\
Cash statistics  & 4.09/4=1.023 & & \\
\hline
\hline
\end{tabular}
%\tablenotetext{a}{The limit magnitude at a significance level of $3\sigma$ is derived from the measurements of the field stars. }
\end{table*}

\begin{figure}[ht!]
\plotone{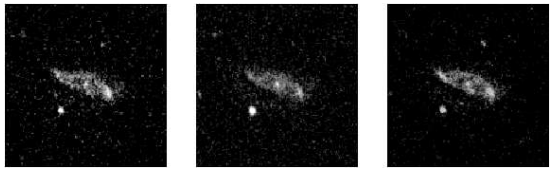}
\caption{From left to right, the co-added \it Swift\rm/UVOT sub-images in the $uvm2$, $uvw1$ and $uvw2$ filters. The size of each image is
70\arcsec$\times$70\arcsec, which corresponds to a physical size of  
$\mathrm{22\mathrm{kpc}\times22\mathrm{kpc}}$ at the redshift. North is in up direction, and east in left direction.
\label{fig:general}}
\end{figure}

\section{Analysis and Results} \label{subsubsec:autonumber}

After the modeling of the new XRT spectrum and the photometries of the new UVOT images,  
we examine the potential relation between the identified spectral type transitions and variations of both 
X-ray and UV emission by comparing the new measurements with previously reported ones .

\subsection{X-ray Emission}

UGC\,3223 was reported as a bright source in the \it ROAST \rm All-sky Survey (RASS, Voges et al. 1999).
The spectrum taken by Stirpe (1990) in 1987 indicates the object was a Seyfert 1.5 galaxy when the RASS was conducted in 1990-1991.
The object is also included in the second Swift-XRT Point Source (2SXP) catalog (Evans et al. 2020) that  
contains 206,335 point sources detected by the \it Swift\rm/XRT in the 0.3-10 keV energy range in the PC mode
between 2005/01/01 and 2018/08/01. For the case of UGC\,3223, the observations used to build the catalog were performed 
between 2013/07/22 and 2014/04/12. The spectroscopy taken by Wang et al. (2020) shows that the object was 
changed to be a Seyfert 2 galaxy with a complete disappearance of its broad Balmer line emission at that time.

Table 3 compares the results extracted from RASS, 2SXP and the new XRT observations performed in 2020. 
Columns (3) and (4) 
list the observed fluxes in 0.1-2.4keV ($F_{\mathrm{0.1-2.4keV}}$) and 0.3-10keV ($F_{\mathrm{0.3-10keV}}$), respectively. For the 2020 observation,
$F_{\mathrm{0.1-2.4keV}}$ is provided by following RASS by assuming a powerlaw with a fixed photon index of $\Gamma=2.3$ associated 
with a Galactic absorption (Zimmermann et al. 2001). 
While, in order to compare the value in 2SXP, the tabulated $F_{\mathrm{0.3-10keV}}$ is obtained from our 
fitting with the model of $wabs*zwabs*powerlaw$, in which both $\Gamma$ and $N_\mathrm{H}$ are free parameters.        
Columns (5) and (6) list the \it ROSAT \rm hardness ratios: 
HR1 and HR2\footnote{The \it ROSAT \rm hardness ratios are defined as the count rates ratios: $\mathrm{HR1}=(B-A)/(B+A)$ and 
$\mathrm{HR2}=(D-C)/(D+C)$, where $A$: 0.1-0.4keV, $B$: 0.5-20.keV, $C$: 0.5-0.9keV and $D$: 0.9-2.0keV.}. Columns (7) and 
(8) are the hardness ratios HR1 and HR2\footnote{The \it Swift \rm hardness ratios HR1 and HR2 are 
defined as: $\mathrm{HR1}=(M-S)/(M+S)$ and $\mathrm{HR2}=(H-M)/(H+M)$, where $S$, $M$, and $H$ are the count rates in 0.3-1keV, 1-2keV and 2-10keV energy bands 
respectively.} defined in Evans et al. (2020) for the \it Swift \rm mission.
All the uncertainties shown in the table correspond to a 90\% significant level, after taking into account the proper error
propagation.

Figure 3 shows the long term variation of the X-ray flux of the object, which is accompanied with its spectral type transitions. 
One can see from the figure that the X-ray flux tightly follows the optical spectral transition. 
The closer the spectral type is to Type I, the higher the X-ray flux. 
%A spectral type closer to Type I, a higher X-ray flux will be.   
Compared to the previous \it Swift \rm observations
carried out in 2013-2014 when the object was classified as a Seyfert 2 galaxy, the 0.3-10keV flux obtained from the new 2020 observation increased by an 
order of magnitude when the object returns to a Seyfert 1.8 galaxy. The 0.1-2.4keV flux of the new 2020 \it Swift \rm observation is still lower than 
that of \it ROSAT \rm by more than one order of 
magnitude when the object shows a Seyfert 1.5 spectrum with an evident broad H$\beta$ emission line (Stirpe 1990).

In contrast to the X-ray-brightness-dependent spectral type transitions, the calculated hardness ratios are found to be constant in both 
``turn-on'' and ``turn-off'' spectral types, which suggests an invariable X-ray spectral shape during the CL phenomenon.    

\begin{table*}[h!]
\renewcommand{\thetable}{\arabic{table}}
\centering
\footnotesize
\caption{Comparison of X-ray emission}
\label{tab:decimal}
\begin{tabular}{cccccccccc}
\tablewidth{0pt}
\hline
\hline
Mission & Date of Obs. & $F_{\mathrm{0.1-2.4keV}}$ & $F_{\mathrm{0.3-10keV}}$ & $HR1_{\mathrm{RASS}}$ & $HR2_{\mathrm{RASS}}$ & $HR1_{\mathrm{Swift}}$ &  $HR2_{\mathrm{Swift}}$ & Sp. type & Reference\\
        &              & \multicolumn{2}{c}{$10^{-12} \mathrm{erg\ s^{-1}}$} & &  & & & & \\ 
(1) & (2) & (3) & (4) & (5) & (6) & (7) & (8)& (9) & (10)\\
\hline
\it ROSAT \rm & 1990-1991 & 15.50 & \dotfill & $0.93\pm0.03$ &  $0.28\pm0.07$ & \dotfill & \dotfill & S1.5 & 1\\
\it Swift \rm & 2013-2014 & \dotfill & $0.34\pm0.08$  & \dotfill & \dotfill & $0.295^{+0.400}_{-0.300}$ &  $0.350^{+0.262}_{-0.204}$ & S2 & 2\\
\it Swift \rm & 2020.04   & $0.86^{+0.16}_{-0.14}$     & $3.86^{+1.36}_{-1.00}$  & $0.953\pm0.233$ & $0.664\pm0.204$ &  $0.357\pm0.170$  & $0.351\pm0.117$ & S1.8 & this work\\
\hline
\end{tabular}
\tablecomments{References: 1: Zimmermann et al. (2001); 2: Evans et al. (2020).}
%\tablenotetext{a}{The data are quoted from }
\end{table*}

\begin{figure}[ht!]
\plotone{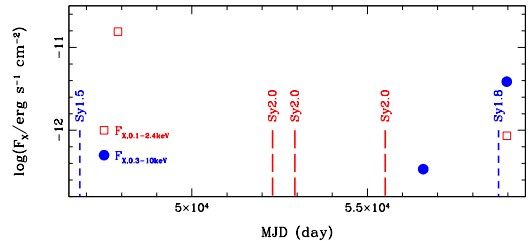}
\caption{Long term variation of X-ray flux of UGC\,3223.  The vertical dashed lines mark the epochs of spectroscopy, 
where the blue short-dashed and red long-dashed lines denote a spectrum with and without broad Balmer lines, respectively.
The corresponding optical spectral types are marked at the end of the lines.
\label{fig:general}}
\end{figure}

\subsection{UV Emission} \label{subsubsec:hide}

Bai et al. (2015) reported an integrated \it GALEX \rm D25 ellipse near-ultraviolet (NUV) magnitude of $16.17\pm0.02$ mag (AB) for the object, 
where only the statistical uncertainties are included.
Among the three \it Swift\rm/UVOT filters, the $uvm2$ filter is most tightly related with the \it GALEX \rm NUV filter with a 
color correction of only 0.013 mag. Taking into account of the systematic zero point uncertainty of NUV of 0.15 mag, the 
measured integrated brightness of $uvm2=16.31\pm0.04$ suggests a consistence with the \it GALEX \rm result, although the object was observed by
\it GALEX \rm by 2013. We argue that the consistence is not hard to understand since the UV emission of the object is found to be 
dominated by the outer host disk. A crude aperture photometry with an aperture size of 5\arcsec\ at the center of the object returns a nuclear
brightness of $uvm2=18.75\pm0.09$ mag, which suggests that the UV emission from the nucleus is only 10\% of the total emission.

\section{Conclusion and Discussion}

Based on the new \it Swift\rm/XRT observations of the nearby repeat CL-AGN UGC\,3223, a comparison with previous observations 
enables us to identify a linkage between X-ray luminosity level and its optical spectral type. 
The object shows the lowest X-ray luminosity level when it was at the ``turn-off'' state with a Seyfert 2 spectrum.
While the highest (moderate) X-ray luminosity level is found when the object was a Seyfert 1.5 (1.8) galaxy.
An invariable X-ray spectral shape is, however, found in the spectral type transitions due to its invariable X-ray hardness ratios. 
In addition, the new
\it Swift \rm/UVOT observations indicate that a fraction of $\sim90$\% UV emission is contributed from 
an intensive star formation activity occurring in the disk of the host galaxy. No variation of its integrated brightness in NUV is
detected during the optical type transitions, which is likely an observational bias due to the low contrast between the host and nucleus.

\subsection{Star Formation Contribution}

We here demonstrate that the X-ray emission contributed by the host star formation activity has negligible effect on our study. 
It is known that star formation activity is also an X-ray emitter due to the collective emission of the XRB population 
(see review in Fabbiano 2006). The X-ray emission can be used as a good tracer of recent star formation activity because the 
evolutionary time-scale of high-mass X-ray binaries (HMXBs) is not longer than $\sim10^7$ yr 
(e.g., Bauer et al. 2002; Grimm et al. 2003; Ranalli et al. 2003; Persic \& Rephaeli 2007; 
Lehmer et al. 2010). 
At the beginning, we estimate the star formation rate (SFR) from its NUV and infrared\footnote{The object is an IRAS bright source: IRAS\,04565+0454.} luminosities basing upon the calibration 
of $\mathrm{SFR}=(L_{\mathrm{NUV}}+L_{\mathrm{IR,24\mu m}})/10^{43.17}\ \mathrm{M_\odot yr^{-1}}$ (Hao et al. 2011), which yields an estimation of 
$\mathrm{SFR\simeq2.4 M_\odot yr^{-1}}$. Given the relationship of $L_\mathrm{X,0.5-8keV}\approx2.6\times10^{39}(\mathrm{SFR}/M_\odot\ \mathrm{yr^{-1}})\ 
\mathrm{erg\ s^{-1}}$ (Mineo et al. 2012), a $L_{\mathrm{X,2-10keV}}(\mathrm{SFR})\simeq3.1\times10^{39}\ \mathrm{erg\ s^{-1}}$ is predicted 
by assuming a powerlaw with $\Gamma=2.0$. This predicted luminosity is far lower than the observed lowest X-ray emission level by almost 2 orders of magnitude.

\subsection{Physics of CL Phenomenon in UGC\,3223}

Taking into account the revealed major role of accretion luminosity on the optical spectral type transitions, we propose here that  
the CL Phenomenon in UGC\,3223 can be understood by two effects, i.e., either a luminosity-dependent BLR or 
a luminosity-dependent dust-torus obscuration, though we can not distinguish which one has more (or equal) impact 
 on the CL Phenomenon at current stage. 

\bf Luminosity-dependent BLR. \rm
With the multi-epoch observations in X-ray, we argue that the observed optical type transitions in UGC\,3223 can be well explained by 
the disk-wind BLR model proposed in Nicastro (2000). Based on a critical radius of the accretion disk
where the power deposited into the vertical outflow is the maximum, a critical $L/L_{\mathrm{Edd}}$ is predicted for disappearance of 
the BLR. 
The unaborbed X-ray luminosity in 2-10keV is estimated to be $L_{\mathrm{X,2-10keV}}=1.4\times10^{42}\ \mathrm{erg\ s^{-1}}$ for the 
2020 \it Swift/\rm XRT observations. 
Adopting a bolometric correction of $L_{\mathrm{bol}}=16L_{\mathrm{X}}$ and a BH mass of $M_{\mathrm{BH}}=9.8\times10^{7}M_\odot$ yields 
an Eddington ratio of $L/L_{\mathrm{Edd}}\simeq0.002$, which is highly consistent with the value estimated from the broad H$\alpha$ line emission
(Wang et al. 2020). Given the variation of its X-ray flux and the invariable hardness ratios, a $L/L_{\mathrm{Edd}}\sim2\times10^{-4}$ is predicted 
when the object shows a Seyfert 2 spectrum with a complete disappearance of its broad Balmer emission lines. 
This $L/L_{\mathrm{Edd}}$ value is clearly lower than the critical one of $2-3\times10^{-3}$ predicted in the disk-wind model (Nicastro 2000) 
by an order of magnitude.

\bf Luminosity-dependent Obscuration. \rm
Although the classical obscuring scenario has been excluded by Sheng et al. (2017, and see also in Ai et al. (2020)) by considering the fact that the 
crossing time of the torus is much larger than the variation time scale of the mid-infrared emission, 
a luminosity-dependent obscuring scenario has been proposed to explain some observed CL cases (e.g., Kishimoto et al. 2013; Oknyansky et al. 2017, 2019, 2020).
It has been known for a long time that the inner most radius of the torus is determined by the hottest dust temperature that is limited 
by sublimation of the dust (i.e., 1500-2000K, e.g., Huffman 1977; Salpeter 1977; Baskin \& Laor 2018). 
A more luminous AGN is therefore predicted to have a larger inner most radius of the torus, which leads to a simple global relation of $r\propto L^{0.5}$ 
between the inner radius $r$ of the torus and AGN luminosity $L$ (Barvainis 1987, 1992). 
This prediction has been confirmed by myriad near-infrared reverberation 
mapping projects (e.g., Nelson 1996; Glass 2004; Minezaki et al. 2006, 2019; Koshida et al. 2014). 
By depending on the average luminosity rather than the instantaneous one, a breath of the inner most radius of the torus was, in fact, revealed in two nearby CL-AGNs: 
NGC\,1566 and NGC\,4151 (e.g., Kishimoto et al. 2009; Oknyansky et al. 2019). An updated analysis indicates a dust recovery time of 
$\sim6$ yrs for NGC\,4151 (Kishimoto et al. 2013), which is shorter than the observed CL timescale of UGC\,3223 of $\sim$30 years.

% In addtion to this globle radius-luminosity relationship, a breath of $r$ was identified in NGC\,4151 by Koshida et al. (2009), even though the 
% breath deviates from the globel $r-L$ relationship. Adopting the universal radius-luminosity relationship of $r\propto L^{0.5}$ yields 
% a relation of $\tan\sigma\propto L^{-1/2}$ between the torus angular width $\sigma$ and luminosity, in which a fixed torus height is assumed.
% Assuming a canonical anguler width of $\sigma_0\sim30\degr$ in the Seyfert 1.5 state with a luminosity of $L_0$, a $\sigma=72\degr$ and
% a $\sigma=84\degr$ are predicted for the Seyfert 1.8 and Seyfert 2 states, respectively.

\acknowledgments

The authors thank the anonymous referee for his/her careful review and helpful suggestions improving the manuscript. 
The study is supported by the National Natural Science Foundation of China under grant 11773036, and by the
Strategic Pioneer Program on Space Science, Chinese Academy of Sciences, grant Nos. XDA15052600 \& XDA15016500.
The authors thank the support from the National Key Research and Development Project of China (grant No. 2020YFE0202100). 
J.W. is supported by Natural Science Foundation of Guangxi (2018GXNSFGA281007), and by Bagui Young Scholars
Program. The authors thank Dr. S. Komossa for preparing and submitting the \it Swift \rm ToO request, and 
for comments and suggestions in X-ray analysis. 
We thank the \it Swift \rm Acting PI, Brad Cenko, for approving our ToO request, and the \it Swift \rm observation team.   
This study uses the NASA/IPAC Extragalactic
Database (NED), which is operated by the Jet Propulsion
Laboratory, California Institute of Technology.

\vspace{5mm}
\facilities{Swift(XRT and UVOT)}
\software{HEASOFT, XSPEC (Arnaud 1996)}

%\bibliography{sample63}{}

\vskip12pt
       

\begin{thebibliography}{}

\bibitem[Ai et al. (2020)]{Ai20} Ai, Y. L., Dou, L. M., Yang, C. W., et al. 2020, \apjl, 890, 29 
\bibitem[Arnaud (1996)]{arn96} Arnaud, K. A. 1996, ASPC, 101, 17 
\bibitem[Antonucci (1993)]{ant93} Antonucci, R. R. J. 1993, \araa, 31, 473
\bibitem[Bai et al. (2015)]{bai15} Bai, Y., Zou, H., Liu, J. F., \& Wang, S. 2015, \apjs, 220, 6 
\bibitem[Barvainis (1987)]{bar87} Barvainis, R. 1987, \apj, 320, 537
\bibitem[Barvainis (1992)]{bar92} Barvainis, R. 1992, \apj, 400, 502
\bibitem[Baskin \& Laor (2018)]{bal18} Baskin, A., \& Laor, A. 2018, \mnras, 474, 1970
\bibitem[Bauer et al. (2002)]{bau02} Bauer, F. E., Alexander, D. M., Brandt, W. N., Hornschemeier, A. E., Vignali, C., Garmire, G. P., \& Schneider, D. P. 2002, \aj, 124, 2351 
\bibitem[Cardelli et al. (1989)]{car89} Cardelli, J. A., Clayton, G. C., \& Mathis, J. S. 1989, \apj, 345, 245
\bibitem[Cash (1979)]{cas79} Cash, W. 1979, \apj, 228, 939
% \bibitem[Cohen et al. (1986)]{coh86} Cohen, R. D., Rudy, R. J., Puetter, R. C., Ake, T. B., \& Foltz, C. B. 1986, \apj, 311, 135 
% \bibitem[Dexter \& Begelamn (2019)]{deb19} Dexter, J., \&  Begelman, M. C. 2019, \mnras, 483, L17  
% \bibitem[Denney et al. (2014)] {den14} Denney, K. D., De Rosa, G., Croxall, K., et al. 2014, \apj, 796, 134
\bibitem[Courtois et al. (2009)]{cou09} Courtois, H. M., Tully, R. B., Fisher, J. R., Bonhomme, N., Zavodny, M., \& Barnes, A. 2009, \aj, 138, 1938
\bibitem[Deo et al. (2006)]{deo06} Deo, R. P., Crenshaw, D. M., \& Kraemer, S. B. 2006, \aj, 132, 321
\bibitem[Elitzur \& Ho (2009)]{elh09} Elitzur, M., \& Ho, L. C. 2009, \apjl, 701, 91
\bibitem[Elitzur \& Shlosman (2006)]{els06} Elitzur, M., \& Shlosman, I. 2006,  \apjl, 648, 101
\bibitem[Evans et al. (2020)]{eva20} Evans, P. A., Page, K. L., Osborne, J. P., et al. 2020, \apjs, 247, 54
\bibitem[Fabbiano (2006)]{fab06} Fabbiano, G. 2006, \araa, 44, 323 
\bibitem[Frederick et al. (2019)]{fre19} Frederick, S., Gezari, S., Graham, M. J., et al. 2019, \apj, 883, 31
\bibitem[Gehrels et al. (2004)]{geh04} Gehrels, N., Chincarini, G., Giommi, P., et al. 2004, \apj, 611, 1005  
\bibitem[Gezari et al. (2017)]{gez17} Gezari, S., Hung, T., Cenko, S. B., et al. 2017, \apj, 835, 144
\bibitem[Giovanelli et al. (1994)]{gio94} Giovanelli, R., Haynes, M. P., Salzer, J. J., Wegner, G., da Costa, L. N., \& Freudling, W. 1994, \aj, 107, 2036
\bibitem[Glass (2004)]{gla04} Glass, I. S. 2004, \mnras, 350, 1049
\bibitem[Guo et al. (2019)]{guo19} Guo, H. X., Sun, M. Y., Liu, X., Wang, T. G., Kong, M. Z., Wang, S., Sheng, Z. F., \& He, Z. C. 2019, \apjl, 833, 44 
\bibitem[Graham et al. (2020)]{gra20} Graham, M. J., Ross, N. P., Stern, D., et al. 2020, \mnras, 491, 4925
\bibitem[Grimm et al. (2003)]{gri03} Grimm, H. -J., Gilfanov, M., \& Sunyaev, R. 2003, \mnras, 339, 793
\bibitem[Hao et al. (2011)]{hao11} Hao, C. -N., Kennicutt, R. C., Johnson, B. D., Calzetti, D., Dale, D. A., \& Moustakas, J. 2011, \apj, 741, 124 
% \bibitem[Graham et al. (2019)]{gra19} Graham, M. J., Ross, N. P., Stern, D., et al. 2020, \mnras, 491, 4925
% \bibitem[Greene \& Ho (2005)]{grh05} Greene, J. E., \& Ho, L. C. 2005, \apj, 630 ,122
% \bibitem[Greene \& Ho (2007)]{grh07} Greene, J. E., \& Ho, L. C. 2007, \apj, 670, 92
% \bibitem[Husemann et al. (2016)]{hus16} Husemann, B., Urrutia, T., Tremblay, G. R., et al. 2016, \aap, 593, L9
\bibitem[Huffman (1977)]{huf77} Huffman, D. R. 1977, AdPhy, 26, 129
\bibitem[Humphrey et al. (2009)]{hum09} Humphrey, P. J., Liu, W. H., \& Buote, D. A. 2009, \apj, 693, 822  
\bibitem[Hutsemekers et al. (2019)]{hut19} Hutsemekers D., Agis Gonzalez B., Marin F., Sluse D., Ramos Almeida C., Acosta Pulido J.-A., 2019, \aap, 625, A54
% \bibitem[Jiang et al. (2016)]{jia16} Jiang, Y. F., Davis, S. W., \& Stone, J. M. 2016, \apj, 827, 10
\bibitem[Kalberla et al. (2005)]{kal05} Kalberla, P. M. W., Burton, W. B., Hartmann, D., Arnal, E. M., Bajaja, E., Morras, R., \& Poppel, W. G. L. 2005, \aap, 440, 775 
\bibitem[Kaastra (2017)]{kaa17} Kaastra, J. S. 2017, \aap, 605, 51  
\bibitem[Kishimoto et al. (2013)]{kis03} Kishimoto, M., Honig, S. F., Antonucci, R., et al. 2013, \apjl, 775, 36
\bibitem[Kishimoto et al. (2009)]{kis09} Kishimoto, M., Honig, S. F., Tristram, K. R. W., \& Weigelt, G. 2009, \aap, 493, L57
\bibitem[Kollatschny et al. (2020)]{kol20} Kollatschny, W., Grupe, D., Parker, M. L., et al. 2020, \aap, 638, 91
\bibitem[Kollatschny et al. (2018)]{kol18} Kollatschny, W., Ochmann, M. W., Zetzl, M., Haas, M., Chelouche, D., Kaspi, S., Pozo Nuñez, F., \& Grupe, D. 2018, \aap, 619, 168 
\bibitem[Koshida et al. (2014)]{kos14} Koshida, S., Minezaki, T., Yoshii, Y., et al. 2014, \apj, 788, 159 
\bibitem[LaMassa et al. (2015)]{lam15} LaMassa, S. M., Cales, S., Moran, E. C., et al. 2015, \apj, 800, 144
\bibitem[Lawrence (2018)]{law18} Lawrence, A. 2018, \nat\ Astronomy, 2, 102
\bibitem[Lehmer et al. (2010)]{leh10} Lehmer, B. D., Alexander, D. M., Bauer, F. E., Brandt, W. N., Goulding, A. D., Jenkins, L. P., Ptak, A., \& Roberts, T. P. 2010, \apjl, 724, 559  
\bibitem[MacLeod et al. (2019)]{mac19} MacLeod, C. L., Green, P. J., Anderson, S. F., et al. 2019, \apj, 874, 8
\bibitem[MacLeod et al. (2010)]{mac10} MacLeod, C. L., Ivezic, Z., Kochanek, C. S., et al. 2010, \apj, 721, 1014 
\bibitem[MacLeod et al. (2016)]{mac16} MacLeod, C. L., Ross, N. P., Lawrence, A., et al. 2016, \mnras, 457, 389
% \bibitem[Mainzer et al. (2014)]{mai14} Mainzer, A., Bauer, J., Cutri, R. M., et al. 2014, \apj, 792, 30
% \bibitem[Marin et al. (2019)]{mar19} Marin, F., Hutsemekers, D., \& Agis Gonzalez, B. 2019, proceedings of the 2019's annual conference of the SF2A , arXiv/astro-ph:1909.02801
% \bibitem[Mathur et al. (2018)]{mat18} Mathur, S., Denney, K. D., Gupta, A., et al. 2018, \apj, 886, 123
% \bibitem[Massey et al. (1988)]{mas88} Massey, P., Strobel, K., Barnes, J. V., et al. 1988, \apj, 328, 315
\bibitem[McElroy et al. (2016)]{mc16} McElroy, R. E., Husemann, B., Croom, S. M., et al. 2016, \aap, 593, L8
\bibitem[Mineo et al. (2012)]{min12} Mineo, S., Gilfanov, M., \& Sunyaev, R. 2012, \mnras, 419, 2095
\bibitem[Minezaki et al. (2006)]{min06} Minezaki, T., Yoshii, Y., Kobayashi, Y., et al. 2006, \apjl, 643, 5 
\bibitem[Nelson (1996)]{nel96} Nelson, B. O. 1996, \apjl, 465, 87 
\bibitem[Nicastro (2000)]{nic00} Nicastro, F. 2000, \apjl, 530, 65
\bibitem[Oknyansky et al. (2017)]{okn17} Oknyansky, V. L., Gaskell, C. M., Huseynov, N. A., et al. 2017, \mnras, 467, 1496 
\bibitem[Oknyansky et al. (2020)]{okn20} Oknyansky, V. L., Winkler, H., Tsygankov, S. S., et al. 2020, \mnras, in print
\bibitem[Oknyansky et al. (2019)]{okn19} Oknyansky, V. L., Winkler, H., Tsygankov, S. S., Lipunov, V. M., Gorbovskoy, E. S., van Wyk, F., Buckley, D. A. H., \& Tyurina, N. V. 2019, \mnras, 483, 558 
\bibitem[Parker et al. (2016)]{par16} Parker, M. L., Komossa, S., Kollatschny, W., et al. 2016, \mnras, 461, 1927
\bibitem[Persic \& Rephaeli (2007) ]{per07} Persic, M., \& Rephaeli, Y.  2007, \aap, 463, 481
% \bibitem[Raimundo et al. (2019)]{rai19} Raimundo, S. I., Vestergaard, M., Koay, J. Y., Lawther, D., Casasola, V., \& Peterson, B. M. 2019, \mnras, 486, 123
% \bibitem[Ross et al. (2018)]{ros18} Ross, N. P., Ford, K. E. S., Graham, M., et al. 2018, \mnras, 480, 4468 
\bibitem[Ranalli et al. (2003)]{ran03} Ranalli, P., Comastri, A., \& Setti, G. 2003, \aap, 399, 39 
\bibitem[Raymond \& Smith (1977)]{ras77} Raymond, J. C., \& Smith, B. W. 1977, \apjs, 35, 419 
\bibitem[Roberts (1962)]{rob62} Roberts, M. S. 1962, \aj, 67, 437 
\bibitem[Ruan et al. (2016)]{rua16} Ruan, J. J., Anderson, S. F., Cales, S. L., et al. 2016, \apj, 826, 188
\bibitem[Ruan et al. (2019)]{rua19} Ruan, J. J., Anderson, S. F., Eracleous, M., Green, P. J., Haggard, D., MacLeod, C. L.\, Runnoe, J. C., \& Sobolewska, M. A. 2019, \apj, 883, 76 
% \bibitem[Rumbaugh et al. (2018)]{rum18} Rumbaugh, N., Shen, Y., Morganson, E., et al. 2018, \apj, 854, 160
\bibitem[Runnoe et al. (2016)]{tun16} Runnoe, J. C., Cales, S., Ruan, J. J., et al. 2016, \mnras, 455, 1691
\bibitem[Salpeter (1977)]{sal77} Salpeter, E. E. 1977, \araa, 15, 267
\bibitem[Schlegel et al. (1998)]{sch98} Schlegel, D. J., Finkbeiner, D. P., \& Davis, M. 1998, \apj, 500, 525
\bibitem[Shapovalova et al. (2010)]{sha10} Shapovalova, A. I., Popovic, L. C., Burenkov, A. N., et al. 2010, \aap, 509, 106
\bibitem[Shappee et al. (2014]{sha14} Shappee, B. J., Prieto, J. L., Grupe, D., et al. 2014, \apj, 788, 48
\bibitem[Sheng et al. (2017)]{she17} Sheng, Z., Wang, T., Jiang, N., et al. 2017, \apjl, 846, 7
\bibitem[Sheng et al. (2020)]{she20} Sheng, Z., Wang, T., Jiang, N., et al. 2020, \apj, 889, 46
\bibitem[Stern et al. (2018)]{ste18} Stern, D., McKernan, B., Graham, M. J., et al. 2018, \apj, 864, 27
\bibitem[Stirpe (1990)] {sti90} Stirpe, G. M. 1990, \aaps, 85, 1049 
\bibitem[Trakhtenbrot et al. (2019)]{tra19} Trakhtenbrot, B., Arcavi, I., MacLeod, C. L., et al. 2019, \apj, 883, 94
\bibitem[Voges et al. (1999)]{vog99} Voges, W., Aschenbach, B., Boller, Th., et al. 1999, \aap, 349, 389 
%\bibitem[Wang et al. (2020a)]{wan20b} Wang, J., Komossa, S., Xu, D. W., Wei, J. Y. 2020, ATel13798, 1
\bibitem[Wang et al. (2020)]{wan20} Wang, J., Xu, D. W., Sun, S. S., Feng, Q. C. Li T. R., Xiao, P. F., \& Wei, J. Y. 2020, \aj, 159, 245
\bibitem[Wang et al. (2019)]{wan19} Wang, J., Xu, D. W., Wang, Y., Zhang, J. B., Zheng, J., \& Wei, J. Y. 2019, \apj, 887, 15
\bibitem[Wang et al. (2018)]{wan18} Wang, J., Xu, D. W., \& Wei, J. Y. 2018, \apj, 858, 49 
\bibitem[Yan et al. (2019)]{yan19} Yan, L., Wang, T. G., Jiang, N., et al. 2019, \apj, 874, 44
\bibitem[Yang et al. (2018)]{yan18} Yang, Q., Wu, X. B., Fan, X. H., et al. 2018, \apj, 862, 109
\bibitem[Zimmermann et al. (2001)]{zim01} Zimmermann, H. -U., Boller, T., D\"{o}bereiner, S., \& Pietsch, W. 2001, \aap, 378, 30 



\end{thebibliography}
%\bibliographystyle{aasjournal}          

\end{document}